\documentclass[]{article}

\def\ben{\begin{equation}}
\def\een{\end{equation}}
\def\bea{\begin{eqnarray}}
\def\eea{\end{eqnarray}}
\input amssym.def
\input amssym.tex
\begin{document}

\hfuzz=100pt
\title{Phantom Matter and the Cosmological Constant}
\author{G. W. Gibbons
\\
D.A.M.T.P.,
\\ Cambridge University,
\\ Wilberforce Road,
\\ Cambridge CB3 0WA,
 \\ U.K.}

\maketitle

\begin{abstract}
Motivated by some recent speculative  attempts to model the dark energy,
scalar fields with negative kinetic energy coupled to gravity
without a cosmological
constant are considered. It is shown that in the presence of
 an ordinary  fluid, any solution of the vacuum Einstein
equations with
cosmological constant is a solution provided
$\rho-P={\Lambda \over 4
  \pi G}$. The solutions can be interpreted as a steady state
in which matter or entropy is being continuously created (or
destroyed). The motion of the matter is not determined
by the background Einstein spacetime, many different matter flows
can be found giving rise to the same metric.
 Solutions without ordinary matter are also considered.
Anti-gravitating multi-solutions and
repulsive solutions which can chase
ordinary matter or black holes are exhibited.
These results may also have applications to
 gravity theories with higher derivatives.

\end{abstract}

\section{Introduction}
Desperate times evidently call for desperate measures.
In order to model the observed acceleration of the scale
factor $a(t)$ of our universe Caldwell \cite{Caldwell} and others
\cite{Carroll}
have turned to matter with  negative energy density. In particular
resort has been made to scalar fields with negative kinetic energies
 \cite{Carroll}. Such ghost or phantom scalar fields have not been
seen in cosmology since the days of the old Steady State theory
with its negative kinetic energy creation field $C(x)$ \cite{HoyleNarlikar}.
The authors of that theory showed that the field equations,
essentially
gravity plus a pressure free  fluid and massless scalar with negative
kinetic energy,
admit a de-Sitter
solution
with constant Hubble constant $H$ with
an exponential scale factor $a(t)=e^{Ht}$ and flat spatial cross
sections satisfying the Perfect Cosmological Principle. Of course
de-Sitter spacetime satisfies the Einstein equations with cosmological
constant $\Lambda = 3H^2$. One of the aims of this paper
is to  point out
that, apparently unknown to those and all subsequent
authors, the equations of motion
also admit as solution {\sl any} solution
of the Einstein equations with cosmological constant
\ben
R_{\alpha \beta}= \Lambda g_{\alpha \beta} \label{cosmo}.
\een
Moreover, given any solution of (\ref{cosmo}), there are infinitely
many
solutions  for $C(x)$, and hence for the distribution of matter.
In other words there is a great deal of indeterminateness in the
theory. Nevertheless what we know of solutions of  (\ref{cosmo})
that at late times strongly indicates that
 they will indeed tend locally to a de-Sitter like state,
at least inside the cosmological horizon of any observer.
This is what is called Cosmic Baldness \cite{BoucherGibbons}.
The necessity of formulating
the statement {\sl locally}, is clearly shown by the example
of Taub-NUT-de-Sitter  spacetime  \cite{Ruback}.
The distinction between Cosmic Baldness and Cosmic
No Hair is made in \cite{Hartnoll}. It is one of
asymptotic stability versus uniqueness. As shown in \cite{Hartnoll}
the  uniqueness property certainly fails in higher dimensions.
For further discussion see  \cite{Barrow} and \cite{Bruni}.

\section{k-Essence}
In their attempts to save appearances, a number of  cosmologists have
resorted to the idea of an as yet undetected and unknown tensile  substance
called quintessence whose function is to cause the scale factor of the
universe
to accelerate at late times by violating the strong energy condition.
 One suggestion is \lq k-essence \rq , a
scalar field $C$ with Lagrangian, in its most general form,
$
L= L(C, y)$,
where $y= g
^{\mu \nu}\partial _\mu C \partial _\nu C$.
The energy momentum tensor is
\ben
T_{\mu \nu} =-2  L_y  C_\mu C_\nu + L g_{\mu \nu} \label{stress}.
\een
From  (\ref{stress}) one easily works out the conditions
$L$ such that the weak dominant or strong energy conditions
are full filled.
An simple example which satisfies the weak and dominant energy conditions
but violates the strong energy condition is Sen's tachyon Lagrangian
\ben
L= -V(C) \sqrt {1+ y}.
\een
If $V(C) = {\rm constant}$ we have a Born-Infeld scalar
which behaves like a Chaplygin gas with the exotic equation of state
  $P= - {A \over \rho},\qquad A >0$.  Despite its exotic properties
this is a causal theory,  small disturbances around a background
are never super-luminal. In general,the characteristic cones are given
by the co-metric \cite{Thoughts}
\ben
{\bigl (G ^{-1} \bigr )} ^{\mu \nu}= g^{\mu \nu} -2L_{yy} \partial ^\mu C \partial C^\nu
\een
with inverse
\ben
G_{\mu \nu}= g_{\mu \nu} -{2 L_{yy} \over L_y + 2L_{yy} }.
\een
Ones' confidence is the existence of k-essence is clearly
diminished if small disturbances can travel super-luminally.
This particular problem   does not arise for a phantom field
for which $L={ 1\over 2} y$ and in a sense
\ben
{\bigl (G ^{-1} \bigr ) }^{\mu \nu}=-g^{\mu \nu}.
\label{signature}\een
Purely formally one may interpret (\ref{signature}) as saying that
phantoms or ghosts feel the opposite spacetime signature
from ordinary matter.

A general feature of k-essence models is that the solutions
will in general develop shocks, typically associated with
caustics at which the equations of motion break down and must
be supplemented by additional physical assumptions. Models
exhibiting shocks are thus essentially incomplete, just like
classical general relativity with its spacetime singularities.
Certain k-essence theories however are exceptional
in this regard and do not admit shocks. Among them
is the Born-Infeld theory, and  its cousin Sen's Lagrangian
for the tachyon \cite{Chaplygin}
. Cosmological consequences of caustics have been discussed recently
in \cite{Caustics}. Clearly, one's confidence in k-essence models
being realized in nature
is increased if they are both causal and shock free.
However the later condition may not be essential
if one thinks of them as being effective, rather than fundamental
theories. Both conditions
certainly hold  for Sen's tachyon Lagrangian \cite{Chaplygin, Roller}.
Sadly however, it does not seem to be suitable
to explain the acceleration of the scale factor of the universe
at late times \cite{Thoughts}.

\section{Coupling of phantom fields to a perfect fluid}

For the simplest phantom field one has $L={1\over 2} y$
and the equations of motion
which  we are trying to solve are :
\ben
R_{\mu \nu} -{1 \over 2} R g_{\mu \nu}  = 8 \pi G \Bigl
((\rho +P ) U_\mu U_\nu +Pg_{\mu \nu} - \partial_ \mu C
\partial _\nu C +{1 \over 2} g_{\mu \nu}  g^{\alpha \beta}
\partial _\alpha C \partial _\beta C
\Bigr ) \label{Einstein}.
\een
 We begin by choosing a Gaussian normal coordinate system
\ben
ds ^2 =- dt ^2 + g_{ij} (t, {\bf x}) dx^i dx ^j.
\een
There are infinitely many ways of doing this for any given spacetime.
Now set
\ben
U_\mu = \partial_\mu  t = \delta^0_\mu \label{flow}
\een
and
\ben
C= a t +b,
\een where $a$ and $b$ are  constants. We now impose (\ref{cosmo}) with
\ben
{\Lambda \over 4 \pi G} = \rho -P
\een
\ben
\rho + P= {1 \over 2} a^2.
\een

Some  remarks are in order.
 If one writes the right hand side of (\ref{Einstein})
as
\ben
 8 \pi G \Bigl ( T^{\rm fluid}_{\mu \nu}  + T^{\rm phantom}_{\mu \nu}
 \Bigr ),
\een
then the sum
 $T_{\mu \nu}= T^{\rm fluid}_{\mu \nu}  + T^{\rm phantom}_{\mu \nu}$
is conserved by the Bianchi identity but the individual summands
are not. This was interpreted by the Steady State theorists
as creation of matter by the creation field in order to maintain
a constant matter density.
In fact the construction works
for any equation of state but the  originators of the theory
restricted themselves to the case $P=0$. In the Lagrangian formulation
originally suggested by Pryce, the $C$ field is coupled to
point particles of mass m by
\ben
 -\sum  m\int d \tau - \sum  m\int \partial _ \mu C d x^\mu \label{lagrange}.
\een
The second term in (\ref{lagrange}) does not contribute to
the equation of motion of the particles but gives
an end-point contribution expressing that the 4- momentum of the
point particles comes from the creation field.

More explicitly, the Bianchi identity applied to (\ref{Einstein})
yields
\ben
(\rho +P) \bigl ( {\dot U} ^\nu + { P^{;\nu} \over \rho +P}
 \bigr )=
 \bigl ( h U^\mu \bigr )_{;\mu}  U^\nu
 -C ^\nu \nabla ^2 C, \label{divergence}
\een
where $h=\rho +P$ is the enthalpy density. In our situation
the right hand side of ({\ref{divergence}) vanishes and we find
\ben
a \nabla ^2  C=  (h U) ^\mu _{;\mu } . \label{source}
\een
Thus the source of the creation field is the rate of production
of enthalpy. On might have though that the divergence of the entropy
current should have arisen here. However in the present situation
where $\rho$ and $P$ are constant, the entropy density $s$
is a constant multiple of the  enthalpy density $h$. In fact one may
re-arrange (\ref{divergence}) using the first law of thermodynamics
 to yield
\ben
(\rho +P) \bigl ( {\dot U} ^\nu +  h^{\nu \mu}
 { P_ {;\mu} \over \rho +P }
\bigr )=
 T\bigl ( s U^\mu \bigr )_{;\mu}  U^\nu
 -C ^\nu \nabla ^2 C,
\een
where $h^{\mu \nu}=g^{\mu \nu} + U^\mu U ^\nu$ and $T$ is the temperature.

Physically one can is tempted to regard the solution as an
out of equilibrium self-organized system maintaining itself
in a steady state  through a uniform  dissipation rate.

By (\ref{flow})
the coordinates $t, x^i$ are co-moving with respect
to the fluid, but as stated above, for a given solution
of (\ref{cosmo}) there are many geodesic vorticity free
congruences.However almost all of these coordinate systems
will develop caustics or other types of singularities.
The region of spacetime covered will thus be geodesically incomplete.
The standard example is de-Sitter spacetime $dS_4$.
In its Steady State guise as a $k=0$, FLRW model
the coordinates cover just one half of the full de-Sitter hyperboloid
embedded in ${\Bbb E} ^{4,1}$
to the future of a null hyperplane passing through the origin.
By contrast in its $k=1$ guise, the coordinate system
covers, in a time symmetric fashion  the entire space.
Particles are first destroyed and then created.

As long as the fluid satisfies the dominant
energy condition, $0\le P \le \rho$, the cosmological constant
is non-negative. The limiting case $\rho=P$, stiff matter,
gives a vanishing
cosmological constant. This is because vorticity free stiff matter
is equivalent to an ordinary massless scalar field  with positive
kinetic
energy. This just cancels the contribution of the
phantom or creation field $C(x)$.

\section{Anti-self-gravitating phantoms}

We now consider what happens in the absence of the fluid.
We shall see that negative energy scalars coupled
to gravity can have some strange
properties. For example  there is a static non-singular
solution with a  wormhole, strictly an Einstein-Rosen
bridge,
with vanishing ADM mass \cite{Rasheed}. It is given by
\ben
ds^2= -dt^2 + { l^2 \over 4} \Bigl (1+ {1 \over r^2 }  \Bigr ) ^2
\bigl ( dx^2 + dy ^2 + dz^2 \bigr),
\een
with $r^2=x^2+y^2 +z^2$, and
\ben
C = \tan ^{-1} \Bigl ( {r^2- 1 \over 2r} \Bigr ).
\een

The involution
\ben
r \rightarrow { 1\over r}
\een
interchanges the two sheets  of the wormhole.
Because the metric is ultra-static, $g_{tt}=-1$ there is no event
horizon
and one can see through from one side to the other.

Perhaps even  more bizarrely, because the ghost scalar field gives rise to
repulsive
rather than attractive forces,
one also has static multi-object solutions.
Using the techniques developed in \cite{anti} one makes the ansatz
\ben
ds ^2 = - \exp (2U) dt ^2 + \exp (-2U) \gamma _{ij} dx ^i dx ^j
\een
and finds that
static solutions may be obtained as stationary points of the
dimensionally reduced
action
\ben
 \int \sqrt {\gamma} \Bigl ( R(\gamma) - 2 \gamma ^{ij} \partial _i U
\partial _j U  + 2 \kappa ^2 \gamma ^{ij} \partial C_i \partial _j C
\Bigr)\een
with $\kappa^2  = 4 \pi G$.
It is now  easy to see that solutions exist  of the form
\ben
ds ^2= -\exp e{2U({\bf x})} dt ^2 + \exp ^{-2U({\bf x})} d{\bf x}^2,
\een
with
\ben
\kappa (C-C_{\infty})=U = \sum { -M_i \over |{\bf x}- {\bf x}_i | }.
\label{antigrav}\een
Each  constant $M_i$ can take  either sign.

The occurrence of solutions with negative ADM mass and the consequent
gravitational repulsions can give rise to runaway
accelerating solutions in which a negative mass particle chases a positive
mass particle  as described in the context of general relativity
by Bondi \cite{Bondi}. The positive mass particle can be a black hole
\cite{Chase}. Thus one might  expect if theories with phantoms
are correct,  to find black holes being chased around by
 supernatural  beings.

\section{Higher spin phantom fields}

Once one has lost one's inhibitions about negative energy, there is no
need
to stop at a single scalar field.
 One could consider negative energy tensor fields.
The case of vector fields, which would arise for example
by dimensional reduction of theories with more than one time, was
discussed in \cite{Rasheed}. One could consider $p$-form gauge
fields, and these may possibly
be related to the creation of  $p$-branes. Ghosts are found
in higher derivative theories \cite{Pais}
Second rank symmetric
tensor ghost are found in higher derivative gravity theories
and these have been invoked to provide inflationary solutions (see
\cite{Hawking} for a recent discussion).
The underlying inflationary
 mechanism may well be related to that described in this
paper.

On an even  more unorthodox tack, it is interesting to note that
bi-metric theories  typically exhibit negative energies.
Take as a representative example, Rosen's theory \cite{Rosen, Stoeger}.
This is based in maps $g_{\alpha \beta}(t,{\bf x}) $ from Minkowski spacetime
${\Bbb E} ^{3,1}$ equipped with its standard Lorentzian metric
$\eta_{\mu \nu}$  to the space $ N\equiv GL(4,{\Bbb R} )/SO(3,1)$
 of Lorentzian metrics $g_{\alpha \beta}$. The Lagrangian density
is taken to be

\ben
{\cal L}= { 1 \over 4}
{\rm Tr } \thinspace \bigl ( g^{-1} \partial_\mu  g^{-1} \partial
^\mu g \bigr ) + { 1\over 8} ( {\rm Tr} \thinspace  g^{-1} \partial g)^2,
\label{Rosen} \een
where the contractions on  the partial derivatives are taken with
respect to the Minkowski metric $\eta_{\mu \nu}$. This is like a
non-linear sigma model except that the field $g_{\alpha \beta}$
transforms like a symmetric tensor under Lorentz-transformations
of the domain
\ben
g (x) \rightarrow L^t g(Lx) L
,\een where $L$ is a Lorentz transformation \footnote{as written this
 is of course a right action}.

In fact the ten- dimensional target space  $N$ is endowed with
an $SL(4,{\Bbb R})$-invariant metric with  signature $(7, 3)$ which is
moreover the product of a one-dimensional factor times a solution
of the  Einstein equations. Kinematically at least, we may think of
the model in which a 3-brane is immersed into a ten-dimensional
spacetime
with three times. To see this explicitly we set $g=\exp \phi U$, where
$\det U =-1$. The target space  metric on $N$
 associated to the action  becomes
\ben
12 d \phi ^2 + {\rm Tr } \thinspace \bigl ( U^{-1} dU U^{-1} dU \bigr )\label{metric} .
\een
The second term in (\ref{metric}) is the standard invariant
metric on the symmetric space $SL(4,{\Bbb R})/SO(3,1)$ induced
from the Killing metric of $SL(4,{\Bbb  R})$ on the totally geodesic
submanifold consisting of metrics invariant under the involution
$g\rightarrow g^t$. The Killing metric on $SL(4,{\Bbb R})$ has
signature $(9,-6)$, the $-6$ coming from the maximal compact subgroup
$SO(4)$. The Killing metric on the Lorentz group has signature
$(3,3)$, the positive directions being the non-compact boosts
and the negative directions the maximal compact subgroup $SO(3)$.
The result is  that the  metric (\ref{metric})  has  signature
$(7,3)$, the 3 negative signs coming  from the components $dg_{0i}$.

Thus, taking into account my signature convention $(-++++)$
  Rosen's  Lagrangian (\ref{Rosen})  has 7 negative energy
excitations . Because   simple solutions exist  of a wavelike character,
the metric $g_{\alpha \beta}$ can be an arbitrary function of
retarded time $t-x^3$, where we have chosen the wave to be moving
in the $x^3$ direction, it is clear  that  both positive and
negative energy excitations can propagate freely.

What about antigravity?
Using the theory of harmonic maps, it is easy to see
that it admits  static anti-self-gravitating
multi-object solutions of the same  form as (\ref{antigrav}).
More generally, by making a diagonal static ansatz in (\ref{Rosen})
it is clear that there are anisotropic solutions of the form
\ben
g_{\alpha \beta} = {\rm diag} \bigl (- \exp 2 V_0({\bf x})
, \exp2 V_1({\bf x}) , \exp V_2({\bf x}) , \exp 2 V_3 ({\bf x})
\bigr ),
\een
where $V_0({\bf x}) ,V_1({\bf x}) , V_2({\bf x} ) ,V_3 ({\bf x})$
are arbitrary harmonic functions on three dimensional Euclidean space
${\Bbb E} ^3$.

The structure of some variable speed of light theories \cite{Drummond}
is rather
similar to Rosen's theory and they may therefore be vulnerable to the
same problems.

\section{Conclusion}

Personally, I think it is rather
premature to abandon the widespread  prejudice
against
theories with negative energy, but if observations were to confirm
the existence of anti-gravitating phantoms one  might be inclined to
reconsider. That the  idea is not completely ridiculous is
clear  from the example of Lifshitz transitions in condensed matter
physics. The coefficients multiplying the  terms involving
the gradient of an order parameter are temperature dependent
and can become negative in a certain temperature range.
This is a disaster, it merely signals that configurations
in which the order parameter varies spatially are thermodynamically
favoured  over configurations in which the order parameter is
spatially homogeneous. Another exotic phenomenon
encountered in low temperature physics is a dispersion
relation for rotons in liquid helium allowing
momentum and velocity to be anti-parallel. Naively
this means accepting negative masses. In both these cases,
the fundamental underlying theory has positive energy.
It is just that one is dealing with fluctuations around a situation
which is not a global equilibrium state.
In any event, it is important to understand what one is
letting oneself in for, if one does contemplate scalars with
negative kinetic
energy.

 \section{Acknowlegement} I would like to thank John Barrow
for helpful conversations.

\end{document}